\documentclass[aps,prl,preprint,groupedaddress,showpacs]{revtex4}

\usepackage{textcomp}

\begin{document}


\title{All-Optical Realization of an Atom Laser}


\author{Giovanni Cennini}
\author{Gunnar Ritt}
\author{Carsten Geckeler}
\author{Martin Weitz}
\affiliation{Physikalisches Institut der Universit\"{a}t T\"{u}bingen, Auf der
Morgenstelle 14, 72076 T\"{u}bingen, Germany}


\date{\today}

\begin{abstract}
  We demonstrate an atom laser using all-optical techniques. A Bose-Einstein
  condensate of rubidium atoms is created by direct evaporative cooling in a
  quasistatic dipole trap realized with a single, tightly focused
  CO$_{2}$-laser beam. An applied magnetic field gradient allows formation of
  the condensate in a field-insensitive $m_{F} = 0$ spin projection only,
  which suppresses fluctuations of the chemical potential from stray magnetic
  fields. A collimated and monoenergetic beam of atoms is extracted from the
  Bose-Einstein condensate by continuously lowering the dipole trapping
  potential in a controlled way to form a novel type of atom laser.
\end{abstract}


\maketitle

The development of optical lasers has revolutionized the field of
light optics. To date, evaporative cooling of bosonic atoms in
traps has allowed for the production of Bose-Einstein condensates
[1]. In magnetic traps, atoms from the macroscopically populated
ground state have been extracted with an applied radio frequency
forming a coherent beam, and this source is commonly referred to
as an atom laser [2-5]. Atom lasers realized so far have all
involved magnetic field sensitive states, where a stray magnetic
field $\Delta B$ causes a shift of the chemical potential of order
$\mu_{B}\cdot \Delta B/k_{\text{B}}$ ($ \cong \Delta B \cdot 67$~nK/mG). It has
been demonstrated that by using magnetic shielding, a
quasicontinuous atom laser beam can nevertheless be successfully
generated in a typical, magnetically noisy lab environment [5]. On
the other hand, the extremely successful development of atomic
frequency standards and atom interferometers has demonstrated the
benefit of using first-order field insensitive ($m_{F} = 0$)
Zeeman states [6]. While atoms in such states cannot be trapped in
magnetic traps, optical dipole force traps can confine atoms in
all spin states. Successful evaporative cooling to Bose-Einstein
condensation has recently been demonstrated in dipole force traps
[7,8].

We here report on the realization of a quantum degenerate gas
in a trap confining atoms in field-insensitive Zeeman states alone. The
remaining magnetic field sensitivity of the chemical potential is
$0.014$~pK/(mG)$^{2}$, and determined only by the quadratic Zeeman
effect. We have extracted the condensed atoms into a collimated,
monochromatic beam to form an all-optical atom laser. Our
experiment is based on direct evaporative cooling of rubidium
atoms in a strongly focused CO$_{2}$--laser dipole trap. The use
of a single running wave allows for a very stable operation of the
condensate production. Note that previous experiments creating
quantum degeneracy in dipole traps have either required the use of
a more alignment-sensitive crossed dipole trap geometry [7] or
Feshbach resonances [8,9] to enhance the collisional rate. In our
experiment, a magnetic field gradient induces a force that is
larger than the confining force along the weakly confining axis of
the CO$_{2}$-laser beam, and thus effectively removes all atoms in
field sensitive states during the final stage of the evaporation.
With no applied field gradient, a spinor condensate with 
12000~atoms distributed among the $m_{F} = -1$, $0$, and~$1$
states is produced. When the gradient is activated, typically
7000~atoms condense into the $m_{F} = 0$ component alone. We do
not observe losses from spin-changing collisions. Finally, by
smoothly ramping down the trapping potential we have coherently
extracted atoms from the condensate to form a well collimated atom
laser beam. The coupling rate can be varied by adjusting the ramp
time. The duration of the output is limited only by the
number of trapped atoms.

Studies of cooling atoms in optical dipole traps
towards high phase space densities have first been carried out by the Stanford
group with a YAG-laser trap [10]. Researchers have since then investigated
both the limits of laser and evaporative cooling in optical traps [11].
Friebel et al. noted that in quasistatic dipole traps, as can be realized
with a CO$_{2}$-laser, polarization gradient cooling alone can yield
surprisingly high phase space densities [12]. In an impressive experiment,
Chapman and co-workers have then realized a Bose-Einstein condensate by
evaporative cooling of atoms in a crossed CO$_{2}$-laser trap [7].

Our Bose-Einstein condensate is produced in a single, tightly
focused beam ($\lambda \simeq 10.6$~$\mu$m) derived from a
CO$_{2}$-laser. For this mid-infrared radiation, the trapping
potential for atoms is well approximated by $V =
-\alpha_{\text{s}}/2 \left\vert E\right\vert^{2}$, where $E$
denotes the laser electric field and $\alpha_{\text{s}}$ the
static atomic polarizability.  The trapping potential is confining
for all Zeeman levels of the electronic ground state. Evaporative
cooling in an optical trap is most straightforwardly achieved by
lowering the trapping potential [10]. The energetically highest
atoms escape the trap, and the remaining atoms rethermalize to a
Maxwellian distribution with lower temperature. Mandatory for the
success of this cooling technique is a sufficiently high elastic
collision rate $\Gamma_{s} \simeq nv \sigma$, being responsible
for rethermalization, throughout the evaporation ramp. 
Let us discuss the scaling of this rate on experimental
parameters. Assuming that the atoms are well localized in the
trap, the potential is harmonic and the average potential energy
per dimension is $k_{\text{B}} T/2 = \frac{m}{2}(2\pi\nu_{i})^{2}
\langle x_i^2 \rangle$, where $\nu_{i}$ denotes the vibrational
frequency and $\langle x_i^2 \rangle$ the mean quadratic spatial
extension of the cloud respectively along one axis. We
infer that the atomic density at a given number of atoms $N$
scales as $N \nu_{x}\nu_{x}\nu_{z}/T^{3/2}$. The vibrational
frequency is strongly dependent on the beam focus: along the beam
axis $\nu_{z} \propto \lambda\sqrt{P}/w_{0}^{3}$, while
transversally one obtains $\nu_{x,y} \propto \sqrt{P}/w_{0}^{2}$
as a function of beam power $P$, beam waist $w_{0}$ and
laser wavelength $\lambda$. The collisional rate thus scales as
$\Gamma_{s} \propto N P^{3/2}/(Tw_{0}^{7})$. During the course of
evaporation, one usually works at a constant ratio of $\eta =
V/k_{\text{B}}T$ where $V$ ($\propto P$) denotes the trap depth so
that we arrive at $\Gamma_{s} \propto N\sqrt{T}/w_{0}^{7}$. The
number of atoms transferred into the dipole trap from the MOT
depends on the beam focus, but a precise model here requires more
parameters. For the sake of simplicity assume that $N \propto
w_{0}^{2}$, so that a $1/\omega_{0}^{5}$ scaling of the collisional
rate remains. During the course of the evaporation, the
collisional rate in optical traps slows down both due to the
reduction in $T$ and the loss of atoms, as was discussed earlier
[7,10,13]. It is demonstrated here that with a comparatively 
small beam waist Bose-Einstein condensation can be achieved 
in a running beam geometry. Besides the technical simplicity compared to a crossed
beam geometry, the moderate confinement along the beam axis allows
to remove magnetic field sensitive states during the course of the
evaporation by applying comparatively low field gradients.

Our experimental setup is as follows. The near-resonant radiation
for cooling and trapping of atoms is generated by two grating
stabilized diode lasers. A ``cooling laser'' is operated with
variable (red) detuning from the $F = 2 \rightarrow F' = 3$
hyperfine component of the rubidium $D2$-line. Its radiation is
amplified by an injection locked diode. We use a 
``repumping laser'' tuned resonantly to
the $F = 1 \rightarrow F' = 2$ hyperfine component. Both optical
beams are spatially filtered, expanded to a 20~mm beam diameter
and spatially overlapped before being directed to a vacuum
chamber. The total optical power here is 42~mW for the cooling
light and 9~mW for the repumping light. A rf-excited
CO$_{2}$-laser generates radiation near $10.6$~$\mu $m. The
mid-infrared radiation passes an acousto-optic modulator used for
controlling the beam intensity, and then enters the vacuum chamber
through a ZnSe window. An adjustable, spherically corrected lens
($f = 38.1$~mm) placed inside the vacuum chamber focuses the beam,
which is oriented horizontally, to a waist of 27~$\mu $m. A pair
of magnetic coils oriented in Anti-Helmholtz configuration
generate a magnetic quadrupole field with a 10~G/cm field
gradient. This field is used both to operate the MOT and to remove
field-sensitive Zeeman components during evaporative cooling.

A typical experimental run proceeds as follows. In a 30~s long MOT
loading phase, we capture $6\cdot 10^{7}$ atoms ($^{87}$Rb) from
the thermal background gas emitted by heated rubidium
dispensers. During this phase, the cooling laser is operated with
a detuning of 18~MHz to the red of the cooling transition.
Subsequently, we increase the detuning to 160~MHz and
simultaneously decrease the power of the repumping beams to $\sim
1/100$ of its initial value for a period of 60~ms. By this
temporal dark MOT, the atomic cloud is compressed and pumped into the
lower hyperfine state as described in more detail in Refs. 10 and
12. Throughout this cycle the CO$_{2}$-laser beam, whose focus
overlaps with the MOT center, is left on. Note that in contrast to
more closely resonant dipole traps, the atomic polarizability for
a quasistatic field is positive for both the lower and upper state
of the cooling transition. Following the dark MOT phase, the repumping light is extinguished
and after an additional delay of 2~ms also the cooling light. By
this time, the atoms are confined in the quasistatic
trapping laser field alone. To analyze the trapped atomic cloud at
the end of an experimental run, the CO$_{2}$-laser beam is
extinguished and shadow images are recorded after a variable free
expansion time.

In initial experiments, we studied the trapping of atoms in the
quasistatic dipole potential at full trapping laser power (i.e.
with no forced evaporation). For these measurements, the magnetic
quadrupole field was switched off by the end of the dark MOT
phase. The trap vibrational frequencies have been
measured to be 4.8~kHz transversally and 350~Hz in a direction longitudinally to
the beam axis. For very short dipole trapping times, 
atoms not transferred to this 
trap have not yet fallen out of the detection region.
The earliest that we observe a clean signal of 
dipole trapped atoms is 70~ms after the end of the dark-MOT phase, at 
which $4 \cdot 10^{6}$ atoms are detected with a temperature of 
140~$\mu\mathrm{K}$. We observe
subsequent trap loss with a faster than vacuum limited rate, which
we attribute to initial ``natural'' evaporation of atoms. At a total 
dipole trapping time of 100 ms, the number of trapped atoms has 
decreased to one third and the temperature has reduced to  
90~$\mu \mathrm{K}$. The trap loss then slows down and the trap gradually 
reaches its vacuum limited lifetime of 12 seconds.
The atomic density at 100 ms dipole trapping time
is $n \simeq 1.2 \cdot 10^{13}/\text{cm}^{3}$. We derive an impressive
collisional rate of 6.2~kHz and a product 
$n\lambda^{3}_{\text{dB}} \simeq 1.2 \cdot 10^{-4}$. 

To achieve lower atomic temperatures, forced
evaporative cooling is applied. In the dipole trapping phase, the power of
the mid-infrared beam is ramped down to induce a time-dependent
trap potential: $V(t) \propto (1+t/\tau)^{-\beta}$. This function
has been designed to maintain a constant value of $\eta =
V/k_{\text{B}}T$ [13]. Optimum cooling was observed when choosing
values near $\tau =0.45$~s and $\beta = 1.4$. We typically use an
evaporative cooling time of 7~s during which the dipole trapping
beam is reduced to a final power of 200~mW. Fig. 1a shows a
typical time-of-flight shadow image of the atomic cloud recorded
15~ms after extinguishing the CO$_{2}$-laser beam. For this
measurement, the MOT quadrupole field was left on throughout the
experimental cycle. We interpret the image as resulting from an
almost pure Bose-Einstein condensate of 7000 atoms in the $F = 1,
m_{F} = 0$ component of the electronic ground state. Atoms in
field sensitive spin projections are removed during evaporation by
the field gradient, as the dipole force along the beam axis is
comparatively weak [14]. While the trapped atomic cloud is
cigar shaped and elongated along the (horizontal) beam
axis, the aspect ratio is inverted in the shown time-of-flight
image due to the anisotropic release of mean field energy. For
comparison, we have also performed measurements with no applied
magnetic quadrupole magnetic field during the forced evaporation
phase. We then produce a spinor condensate with 12000 atoms
distributed among the $m_{F} = -1, 0, 1$ Zeeman components of the
$F = 1$ ground state. Typical data for this measurement is shown
in Fig. 1b, where a separation into clouds with different spin
projections is clearly visible.

Subsequent measurements have all been carried out with atoms
condensed in the field-insensitive Zeeman component only. We have
analyzed cross sections of shadow images taken for different end
values of the evaporation ramp. The profiles shown in Figs. 2a-2c
illustrate the formation of the condensate from a thermal cloud
in the former to an almost pure Bose-Einstein condensate in the 
latter image. 
We deduce a critical temperature $T_{\text{c}} \simeq
220$~nK and a condensate peak-density $1.2 \cdot
10^{14}/\text{cm}^{3}$. The observed condensate lifetime is
5~s and attributed to be mainly limited by three-body
collisions. A remarkable issue that within this lifetime we do not
observe transfer of atoms into field-sensitive Zeeman states. We
anticipate that spin-changing collisions are
suppressed energetically due to the second order Zeeman effect in
the inhomogeneous field [15].

To outcouple the atoms and form an atom laser beam, after creating
a condensate we further reduce the trapping laser power in a slow
and controlled way. It is clear that the ramp must be smooth
enough to minimize condensate excitations. Experimentally, the
CO$_{2}$-laser beam power is reduced by applying a 100~ms long
linear ramp to the acousto-optic modulator drive. The condensate is
outcoupled once the dipole trapping potential $V(x,y,z) =
-\alpha_{\text{s}}/2 \left\vert E\right\vert^{2} - mgx$ is
approaching the limit of supporting the atomic cloud against
gravity. The duration of coherent output is approximately
$\mu / \frac{\text{d}V_{s}}{\text{d}t}$, where $\mu $ denotes
the chemical potential and $\frac{\text{d}V_{s}}{\text{d}t}$ the
lowering rate of the saddle point of the potential. Fig. 3a shows a
typical image of the outcoupled coherent atom laser beam. We observe an up to 1~mm
long intense, well directed beam of atoms. We expect that the momentum spread is 
limited by the uncertainty principle only. Experimentally, we can
determine the transverse velocity spread along the z-axis to be 
below our resolution of $0.3$~mm/s. The transverse 
distribution of the (averaged) optical column density can be fitted well when 
assuming an inverted parabolic profile for the transversal mode of the beam, 
as shown in Fig. 3b. This form of trial function is used, since in 
the Thomas-Fermi limit the condensate density profile images 
the form of the trapping potential. Of relevance for the transverse 
mode are the saddle point values over the aperture of the atom laser. 
The beam is emitted from the (downward directed) side of the 
condensate. Note that atom lasers based on rf output coupling 
are often operated in a regime where spin flips into untrapped states
are induced in regions interior to  
the condensate [16]. Compared to such a situation, mean field effects 
from residing atoms, which can act as a lens for the emitted beam 
and lead to a transverse interference structure
[17], are reduced in the present scheme. We have attempted to estimate 
the brightness of our atom laser. For the beam shown
in Fig. 3a, the atomic flux is $8.4 \cdot 10^{5} $atoms/s. 
If we assume uncertainty limited transverse velocity spreads ($\Delta
v_y \simeq 0.33$~mm/s and $\Delta v_z \simeq 0.023$~mm/s) and a
Fourier-limited width along the atom laser beam axis ($\Delta v_x
\simeq 0.35$~mm/s), we estimate a brightness of $7 \cdot
10^{27}$~atoms~$\mathrm{s}^2\,\mathrm{m}^{-5}$.  This is orders of
magnitude above results achieved with thermal sources, but
a factor 6 below a similarly derived value for a `conventional' atom
laser given in Ref.~5. Note that our presently used low power diode
cooling lasers yield a comparatively small number of atoms in the MOT,
which will be improved in the future.

As fluctuations of the magnetic
field cause a shift of the chemical potential via the second order
Zeeman effect only (yielding an expected shift $(\mu_{\text{B}}
\cdot \Delta B)^{2}/(E_{\text{HFS}} \cdot k_{\text{B}})$), no
magnetic shielding is required for this measurement even in a
magnetically noisy environment. It remains important to note that
the stability of an atom laser beam is determined by energy
fluctuations during the outcoupling process associated with 
\textit{both} 
the extracted and the residing atoms. Earlier works, partly
realizing output coupling by transfer from field sensitive into
field insensitive spin projections, thus did experience first-order
sensitivity to field fluctuations [2-5]. The output stability 
of the present scheme is for a smooth ramp limited only
by trapping laser intensity fluctuations.

To conclude, we report on the realization of a Bose-Einstein condensate by
direct evaporative cooling in a single beam CO$_{2}$-laser dipole trap. The
atoms condense into a magnetic field insensitive state. Our results
culminate in the all-optical realization of an atom laser.

For the future, we anticipate that the
demonstrated method to produce Bose-condensed atoms opens up applications
in fundamental and applied sciences. The demonstrated technique could furthermore stimulate
studies of two-component quantum gases driven by exciting the microwave clock
transition between the hyperfine ground states $F = 1, m_{F} = 0$ and $F = 2,
m_{F} = 0$. It clearly would be of importance to in detail study the stability
of different spin projections (and spin mixtures) towards spin-changing
collisions. Finally, we foresee that the generated atom laser beam is awaiting
fascinating applications in atom optics experiments. Future developments here
can allow for improved atom interferometric measurements of gravitation and
rotation.

We acknowledge partial financial support from the Deutsche
Forschungsgemeinschaft, the Landesstiftung Baden W\"urttemberg,
and the European Community.

\textit{Note added:} After submission of this work, Y. Takasu et
al. published an approach to produce field-insensitive
condensates based on atoms with spin singlet ground states [18].

\small{

\newpage
\begin{figure}
\caption{%
  False-color shadow images of the atomic cloud after 15 ms of free expansion (field of
  view: 0.33 mm square). (a) Stern-Gerlach magnetic field gradient applied
  throughout the experiment, so that a pure $m_{F}= 0$ condensate is produced.
  (b) Field gradient activated only during free expansion phase. The three
  spin projections $m_{F}= -1$, $0$, and $1$ of a spinor condensate are visible as
  separate clouds.}
\end{figure}
\begin{figure}
\caption{%
  Vertical cuts through shadow images recorded for pure $m_{F}= 0$ ensembles
  for different final trap laser powers after 15 ms of free expansion.
  (a) 500 mW: yielding a thermal cloud  with $T \simeq 350$~nK, (b) 240 mW: 
  partly condensed ($T \simeq 180$~nK). The
  dashed line is to guide the eye in distinguishing the broad thermal
  background from the narrow condensate peak. (c) 200 mW,
  resulting in an almost pure BEC.}
\end{figure}
\begin{figure}
\begin{minipage}{\linewidth}
\end{minipage}
\caption{%
  (a) False-color shadow image of the generated atom laser beam. 
  The field of view comprises 0.28 mm by 0.5 mm. (b) Fitted transverse 
  cut of the image averaged over a 0.19 mm beam length.}
\end{figure}
\bibliography{basename of .bib file}

\begin{thebibliography}{10}
\bibitem{14}
Experimentally, the minimum of the magnetic quadrupole field is not
necessarily precisely overlapped with the center of the dipole trap. 
Note that in a crossed dipole trap different magnetic gradients 
of magnitude below the threshold of untrapping field-sensitive states 
have been used to alter the magnetisation in a condensate [7], 
though it was noted that the origin of this magnetisation yet
remained to be determined.
\end{thebibliography}

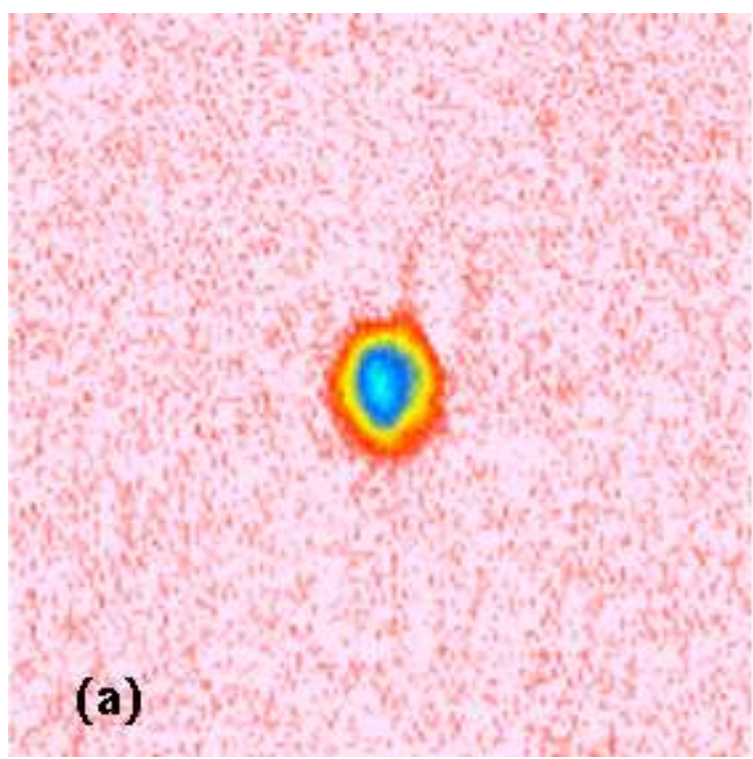
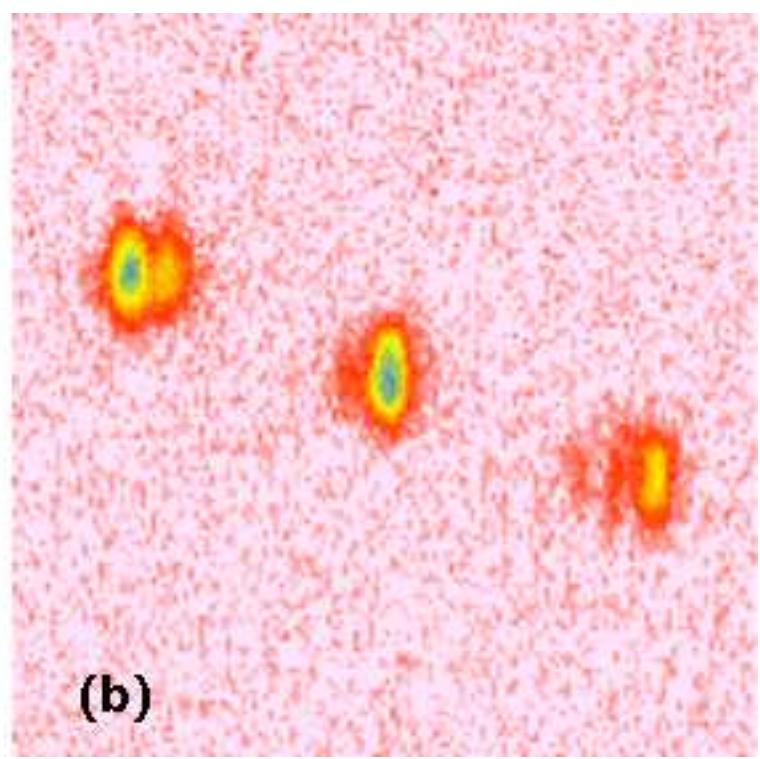
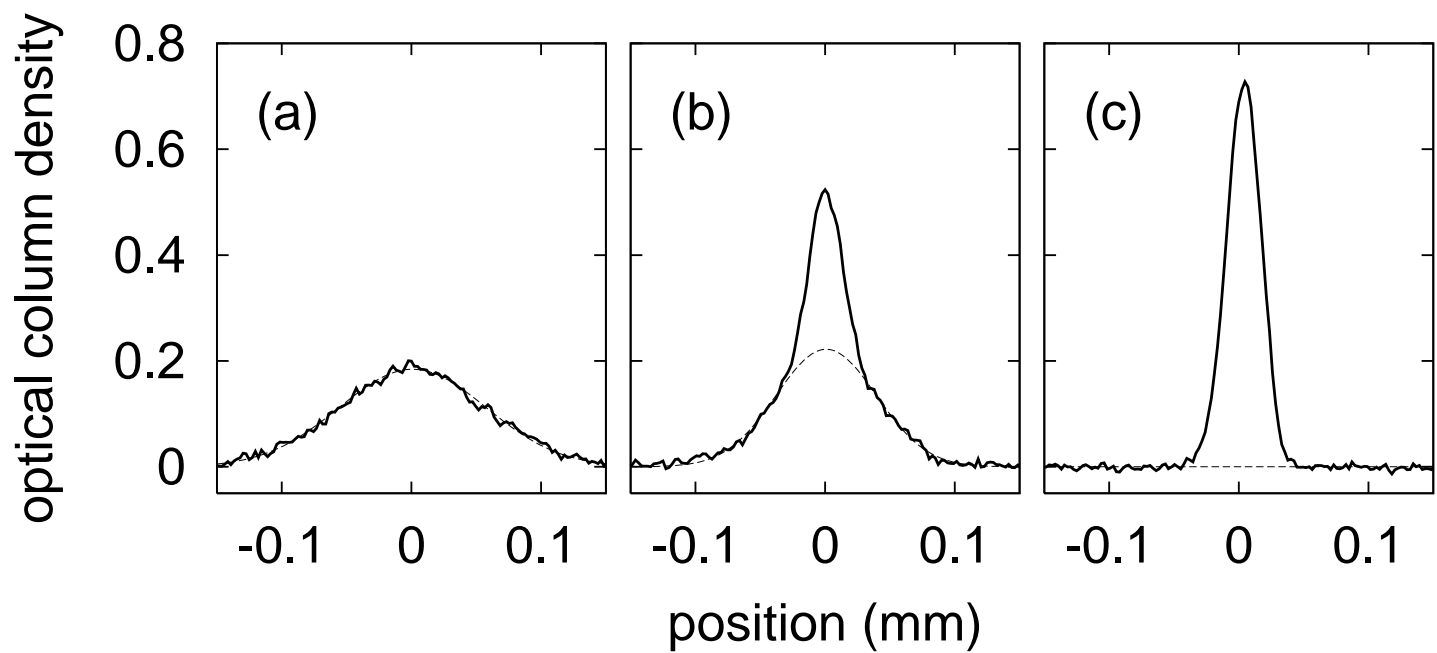
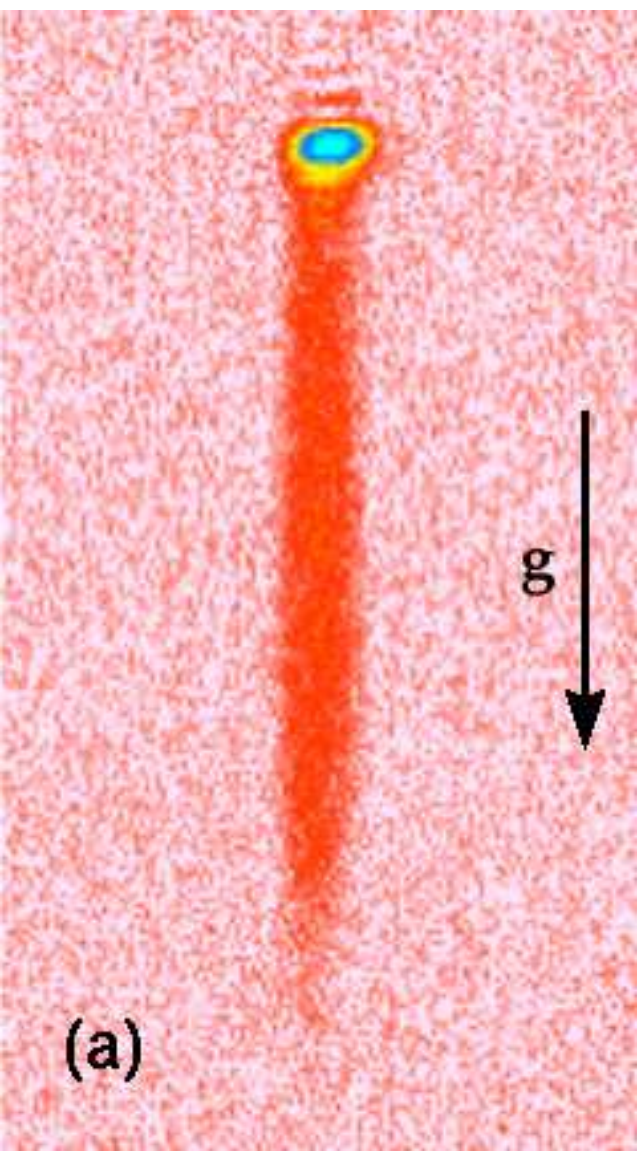
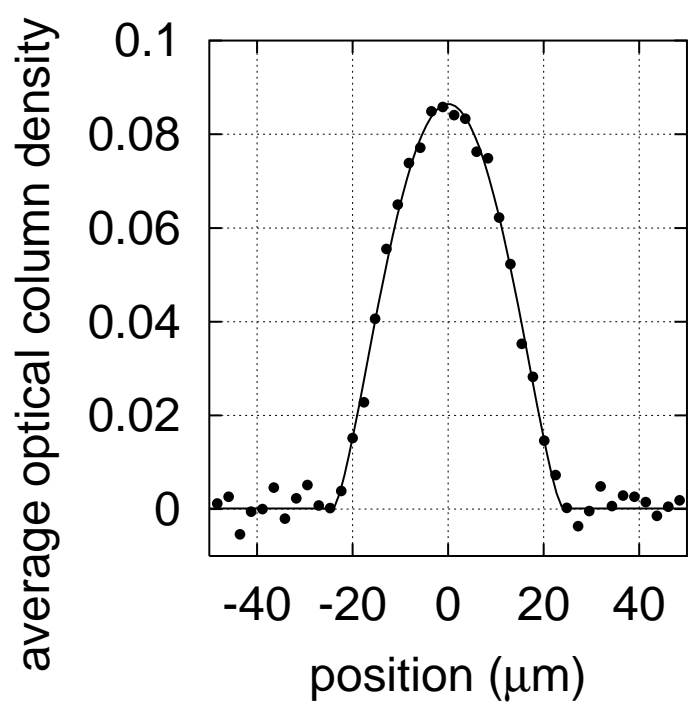
\end{document}